\documentclass[showkeys,aip,graphicx,amsmath,superscriptaddress,amssymb]{revtex4-2}

\usepackage{CJKutf8}
\usepackage{lineno,hyperref}
\usepackage[normalem]{ulem}
\usepackage{graphics}
\usepackage{amsmath,bm}
\usepackage{amssymb}
\usepackage{graphicx}
\usepackage{bm}
\usepackage{amsfonts}
\usepackage{color}
\usepackage{epstopdf}
\usepackage{geometry}
\usepackage{booktabs}
\usepackage{array}
\usepackage{hyperref}
\usepackage{graphics}
\usepackage{epsfig}
\usepackage{algorithm}
\usepackage{algpseudocode}
\usepackage{bm}
\usepackage{epic,eepic}
\usepackage{epsfig}
\usepackage{pifont}
\usepackage{braket}
\usepackage{subfigure}

\usepackage{xspace}
\usepackage{graphicx,color}
\usepackage{amsmath,amssymb,amsfonts,mathrsfs}
\usepackage[normalem]{ulem}
\usepackage{booktabs}
\usepackage{float}
\usepackage[toc,page]{appendix}
\usepackage[capitalize]{cleveref}
\usepackage{mathtools}
\usepackage[normalem]{ulem}
\usepackage{gensymb}
\usepackage{lineno}
\usepackage{hyperref}

\begin{document}

\title{Explainable quantum-compressed machine learning for complex fluid flows}

\author{Xiao Xue$^{\dagger}$}
\affiliation{ 
Centre for Computational Science, University College London, London, UK
}
\author{Maida Wang$^{\dagger}$}
\affiliation{ 
Centre for Computational Science, University College London, London, UK
}
\author{Mingyang Gao}
\affiliation{ 
Centre for Computational Science, University College London, London, UK
}
\author{Minh Chung}
\affiliation{
Leibniz Supercomputing Centre of the Bavarian Academy of Sciences and Humanities,\\
Boltzmannstraße~1, 85748~Garching, Germany
}

\author{Peter V. Coveney}%
\email{p.v.coveney@ucl.ac.uk}
\affiliation{ 
Centre for Computational Science, University College London, London, UK
}
\affiliation{
Centre for Advanced Research Computing, University College London, London, UK
}
\thanks{$^{\dagger}$These authors contributed equally as the first author.}

\date{\today}

\begin{abstract}
Machine-learning surrogates of physical systems face a paradox: explainable models facing the challenge of expressivity to capture complex nonlinear flows, whereas expressive deep surrogates match high-fidelity simulations only through massive parameterisations that turn the learned dynamics into a black box. Here, we introduce quantum-compressed machine learning (QCML), which resolves this tension by compressing the latent propagator of a flow surrogate from $524{,}288$ trainable parameters to no more than $8$. This parameter reduction brings the learned dynamical law to the parameter scale of a physical constitutive relation rather than a black-box neural network, making the surrogate directly interpretable and controllable without sacrificing expressivity. The compression is realised by a structured quantum circuit whose unitary propagator constrains the latent spectrum to the unit circle exactly and by construction, replacing exponential error growth with linear accumulation over autoregressive rollouts. Classical regularisation only approximates this constraint: even a quantum-inspired classical baseline penalised towards unitarity collapses within one Lyapunov time on turbulent channel flow, whereas QCML remains stable over the full rollout. Shared phase and coupling angles parameterising the circuit correspond directly to modal frequencies and inter-mode interactions, giving the learned dynamics a physical interpretation in spectral space. On two patient-specific cardiovascular benchmarks, the structured QCML propagator matches the predictive accuracy of its classical counterpart on surface pressure spectra, pressure drop, and wall shear stress. These results establish QCML as a working component of scientific machine learning and a concrete contribution towards practical quantum advantage in real-world prediction.

\end{abstract}

\maketitle

\section{Main}
Predicting how nonlinear physical systems evolve in time is a central problem of computational science, spanning turbulence in engineering~\cite{duraisamy2019turbulence}, climate dynamics~\cite{bi2023accurate,lam2023learning}, astrophysics~\cite{cho2002simulations,gopakumar2024plasma}, and blood flows in cardiovascular medicine~\cite{feiger2020accelerating,xue2025uncertainty}. High-performance computing (HPC) now resolves such systems at unprecedented fidelity, yet each simulation routinely consumes tens of thousands of HPC node hours per second of physical time, far exceeding most decision-making timescales~\cite{tanade2022analysis,leuprecht2003blood,groen2013analysing,mazzeo2008hemelb}. 
Artificial intelligence for science has matured in response, producing deployable surrogate models for fluid mechanics~\cite{brunton2020machine,xue2026uni}, weather~\cite{price2025probabilistic,pathak2026kilometer} and biomedical applications~\cite{passaro2025boltz,bhati2026integrated,wan2026reliability}. Representative families include physics-informed neural networks~\cite{karniadakis2021physics}, neural operators~\cite{li2020fourier} and broader scientific machine-learning architectures that learn solution maps directly from simulation or observational data~\cite{Karniadakis2025priorsr,raissi2019physics}. 
Quantum computing offers a complementary route: quantum algorithms provide potential speed-ups for linear quantum dynamics, most prominently the Schr\"odinger equation, and related ideas have been extended to broader classes of linear partial differential equations (PDEs)~\cite{Preskill_2018,huang2025vast}. Nonlinear physical flows are substantially harder: quantum evolution is linear and unitary, whereas fluid dynamics involves nonlinear transport, dissipation and multiscale closure. Although quantum algorithms for nonlinear PDEs and computational fluid dynamics (CFD) have been proposed, existing demonstrations remain largely limited to low-dimensional model equations or simplified flows~\cite{lubasch2020variational,sanavio2024lattice,sanavio2025explicit,kocherla2024fully,wawrzyniak2024unitary}. Direct quantum CFD therefore remains a longer-term target for fault-tolerant devices. Across these efforts, the central challenge persists: building surrogates of nonlinear physical flows that are not only reliable and fast but also stable over long prediction horizons, parameter-efficient, and transparent in the way they learn the dynamics.

Existing surrogate-modelling families address parts of this challenge, but none address all of it. Reduced-order models built from physical priors expose interpretable spectral structure and admit principled stability arguments. Representative examples include proper-orthogonal-decomposition projections and classical Koopman and dynamic-mode-decomposition approximations~\cite{mezic2021koopman,Brunton2022Koopman,williams2015data,korda2018convergence}. Their nonlinear capacity is, however, bounded by the chosen basis, which limits their ability to resolve the spatially complex, potentially turbulent multiscale flow structures that govern many real applications. Neural operators and deep learning surrogates, including Fourier neural operators~\cite{li2020fourier} and DeepONet~\cite{lu2021learning}, recover nonlinear behaviour with comparatively little inductive bias~\cite{kovachki2023neural,li2021learning,feiger2020accelerating}. Their accuracy, however, relies on millions to billions of trainable parameters whose individual roles remain opaque, eroding the interpretability needed to trust predictions in safety-critical settings. Autoregressive rollouts compound the problem: without an intrinsic stability mechanism, prediction errors accumulate silently across long time horizons~\cite{mccabe2023towards}. 

The alternative near-term route is hybrid quantum-classical learning, in which quantum processors act not as standalone PDE solvers but as components inside classical scientific machine-learning pipelines~\cite{dunjko2016quantum,ghazi2025quantum}. This route introduces its own design challenges. Variational quantum circuits are shallow enough for present devices, but their usefulness depends critically on the ansatz. Generic hardware-efficient ansatz families can suffer from barren plateaus with exponentially small gradients, while problem-agnostic circuits provide few spectral or physical properties~\cite{mcclean2018barren,cerezo2022challenges,lubasch2020variational}. 
A problem-matched quantum prior, by contrast, can be genuinely informative: quantum-informed machine learning shows that such a prior can guide classical machine-learning systems toward more physical solutions and improve their efficiency and accuracy~\cite{wangxue2026quantum,huang2025vast}. 
In that setting the quantum prior is coupled loosely, shaping training from outside the main learning loop while the latent dynamics of the classical model remains heavily parameterised and effectively a black box, so the end-to-end pipeline is also opaque. What has been missing is an end-to-end quantum-classical loop, compressing the latent dynamics into a small, interpretable set of parameters for real-world physical-flow surrogacy. Across these families, three needs remain unmet at once: stable long-horizon rollout, parameter efficiency and explainable latent dynamics.

\begin{figure}[htbp!]
\centering
\includegraphics[width=0.85\textwidth]{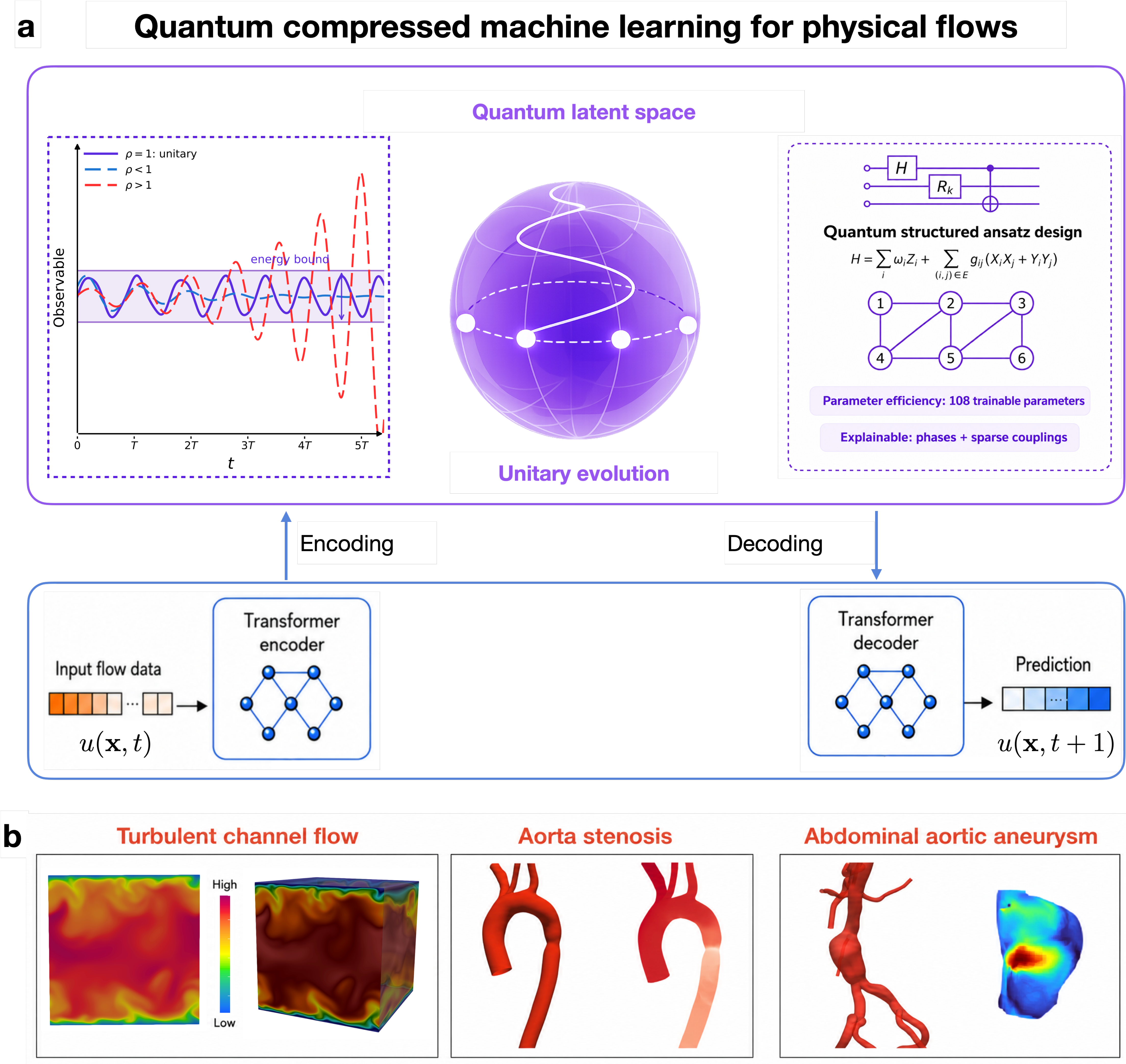}
\caption{\textbf{QCML framework for nonlinear physical flows.}
(\textbf{a}) Hybrid quantum--classical training loop. A transformer encoder maps the input field $u(\mathbf{x},t)$ into a compact latent representation; this latent state is embedded into a quantum Hilbert space, propagated forward by a structured unitary $U_q$, and decoded back into a transformer decoder that returns the next-step field $u(\mathbf{x},t+1)$. Left inset: latent-norm behaviour under different spectral radii $\rho$; $\rho=1$ (unitary) keeps the observable bounded over the rollout, whereas $\rho<1$ and $\rho>1$ produce vanishing and exploding regimes. Centre: Bloch-sphere sketch of the latent state evolving under $U_q$; the equatorial $(x,y)$ plane spans the real and imaginary parts of the off-diagonal coherence and the polar $z$ axis encodes the populations, so $U_q$ traces a norm-preserving trajectory on the unit sphere despite being a complex-valued operator. Right inset: structured ansatz design, showing the Hamiltonian operator that generates $U_q$, the sparse coupling graph, and two design properties: as few as $8$ trainable parameters through parameter sharing, and interpretable phase and coupling parameters.
(\textbf{b}) Three benchmark systems of increasing complexity: 3D turbulent channel flow, stenotic aortic flow, and patient-specific abdominal aortic aneurysm haemodynamics.}
\label{fig:framework}
\end{figure}

Here we introduce quantum-compressed machine learning (QCML), a hybrid quantum-classical framework summarised in Fig.~\ref{fig:framework}a. A classical transformer encoder--decoder~\cite{zamir2022restormer} maps physical states into a compressed quantum latent space, and a structured quantum circuit propagates this latent representation in time. The central design principle is extreme parameter compression: a structured ansatz shares two trainable scalars per layer across all qubits and edges, reducing the latent propagator from ca. $525{,}000$ trainable parameters in the classical baseline to as few as $8$, a $66{,}000$-fold reduction that brings the learned dynamical law to the parameter scale of a physical governing equation rather than a black-box neural network~\cite{Cerezo2021VQA,mitarai2018quantum,goto2021universal,wang2025parameter} (Fig.~\ref{fig:framework}a, top-right inset). This compression delivers the capabilities that existing surrogates lack. Long-horizon stability follows from the unitarity of the quantum propagator, which pins latent eigenvalues to the unit circle and replaces exponential error growth with linear accumulation over autoregressive rollouts~\cite{lyapunov1892general,khalil2002nonlinear,koopman1931hamiltonian,mezic2021koopman,bengio1994,williams2015data,korda2018convergence} (Fig.~\ref{fig:framework}a, top-left inset); this constraint holds as an operator identity for every parameter value, whereas classical surrogates can only approximate it through soft spectral penalties whose residual drift compounds over the rollout (See Methods). The same parameter sharing mitigates the barren-plateau pathology of generic variational circuits~\cite{mcclean2018barren,Cerezo2021CostFunction,Larocca2025BarrenReview} and tightens the generalisation bound~\cite{caro2022generalization}, keeping the model trainable in the small-data regime of physical-flow surrogacy (see Methods). 
Explainability follows from the compression itself: with only a few trainable parameters, each corresponding to an identifiable modal frequency or inter-mode coupling strength, the learned propagator becomes directly readable, unlike the opaque weight matrices of classical surrogates.

We evaluate QCML on three nonlinear systems of increasing complexity: turbulent channel flow inflow, stenotic aortic flow, and patient-specific abdominal aortic aneurysm haemodynamics (Fig.~\ref{fig:framework}b)~\cite{feiger2020accelerating,xue2025uncertainty,tanade2022analysis}. With only $8$ trainable quantum parameters in the structured variant, QCML matches the predictive accuracy of the classical ML baseline on the diagnostically relevant quantities: velocity spectra for turbulence, pressure drop for the stenotic flow, and wall shear stress for the aneurysm. On a noise-aware classical emulator of the structured circuit, per-cardiac-cycle inference is several orders of magnitude cheaper than the ${\sim}3.4\,\mathrm{h}$ required per cycle on $1{,}024$ CPU cores under conventional CFD~\cite{tanade2022analysis}; we report this as the algorithmic cost of the latent propagation, which excludes the state-preparation, measurement-shot and classical--quantum data-transfer overheads of present quantum hardware (Supplementary~S3.6). Backend validation on IQM's 54-qubit Emerald processor yields $79.69\%$ one-step agreement for turbulent channel flow; the two cardiovascular benchmarks, evaluated on a noise-aware emulator, reach $95.54\%$ and $95.25\%$ agreement respectively (Table~\ref{tab:hardware_validation}). To our knowledge, these results demonstrate the first hybrid quantum-classical surrogate for application-scale nonlinear physical flows, spanning turbulence and patient-specific cardiovascular haemodynamics, and point to integrated quantum-HPC workflows as a practical near-term route to broader deployment.

\section{Turbulent channel flow inflow generation}\label{sec:turbulent}

Turbulent channel flow at $Re_\tau = 180$ is a canonical chaotic benchmark for extended autoregressive rollouts~\cite{fukami2019synthetic,fukami2019super,xue2026uni,xue2022synthetic}. We use it to test whether the unitary latent dynamics of QCML translate into stable long-horizon prediction. Rollout time is reported in dimensionless form $t^{*} = t/T_\lambda$, with $T_\lambda$ the Lyapunov time of the resolved ground-truth field (Supplementary~S2.1). Two QCML variants are reported throughout. Both share the same transformer encoder--decoder backbone and act on $N_q=9$ qubits through $L$ variational layers, but differ in the quantum ansatz. The structured variant QCML (S) uses a structured circuit in which each layer applies a shared mode-wise phase rotation on every qubit followed by a shared coupling on every edge of a sparse graph, leaving only $2L$ trainable scalars per circuit (Methods, Eq.~\eqref{eq:Uq_circuit}). The baseline variant QCML uses a generic hardware-efficient circuit, with an independent general single-qubit rotation on every qubit and a fixed entangling ring in each layer, giving $3LN_q$ trainable parameters and no trainable couplings. Comparing the two isolates the effect of imposing Koopman-inspired structure rather than merely reducing the parameter count. The detailed gate-level definitions and per-case counts are given in Supplementary~S3.4 and S4.

Figure~\ref{fig:turbulent}a sketches the simulation that generated the training, validation and test data: a streamwise-driven 3D turbulent channel flow with no-slip top and bottom walls and periodic streamwise/spanwise boundaries~\cite{latt2008straight}. Once the flow had reached a statistically stationary turbulent regime, 320 two-dimensional trajectories of wall-normal velocity were recorded on the mid-plane at $192\times 192$ resolution and downsampled to $64\times 64$; the full data-generation protocol is given in SI~S2.1. The 320 realisations were partitioned 80/10/10 into training, validation and test sets. The classical ML baseline~\cite{wangxue2026quantum} was trained on the same split for comparison.

All surrogates were evaluated under an autoregressive window-rollout protocol $\hat{\mathbf{u}}_{t+1:t+k} = f_{\theta}(\mathbf{u}_{t-k+1:t})$, in which a window of $k$ consecutive snapshots of the predicted field $\mathbf{u}$ is mapped to the next $k$ and the model is then advanced on its own predictions without further reference input (Supplementary~S3.5). Starting from a reference initial window, the rollout was carried out to five Lyapunov times ($t^{\ast}\approx 5$). Figure~\ref{fig:turbulent}b shows instantaneous velocity fields at five instants along this rollout against the ground-truth large-eddy lattice Boltzmann reference, in which a Smagorinsky closure is embedded into the lattice Boltzmann collision step~\cite{smagorinsky1963general,hou1994lattice,koda2015lattice}. Classical ML lost near-wall streaks already by $t^{\ast}\approx 1$ and relaxed toward a nearly featureless mean-like field for the rest of the rollout. Both QCML variants kept the large-scale streaks and boundary-layer coherence visible in the reference out to $t^{\ast}=5.27$, with QCML (S) marginally cleaner than QCML at late times. Unitarity fixes the latent norm by construction (Methods, Sec.~``Stability against exploding and vanishing gradients''), which sustained coherent fluctuations across the plotted window for both QCML variants but was absent in classical ML.

Spectral and statistical diagnostics over the same turbulent-inflow rollout corroborate this stability; the formal definitions of all metrics reported here and in subsequent figures are collected in SI~S6. Figure~\ref{fig:turbulent}c shows that both QCML variants overlap the reference spectrum $\langle E(k)\rangle$ over more than a decade of wavenumber space, whereas classical ML overshoots the reference at high $k$ and accumulates spurious small-scale energy. In Fig.~\ref{fig:turbulent}d, the mean streamwise velocity in wall units lies on the direct numerical simulation (DNS) reference and the log law at $Re_\tau=180$ for every model, indicating that the wall-normal profile is recovered correctly by both QCML variants and the classical baseline. Figure~\ref{fig:turbulent}e then shows that both QCML variants reproduce the bulk-flow density of the reference around the peak at $u\approx 0.027$ in lattice Boltzmann units (LBU; conversion to physical SI units is detailed in SI~S1), whereas classical ML sharpens this peak and underpopulates the tails, consistent with the loss of small-scale variability already visible in Fig.~\ref{fig:turbulent}b; the agreement is achieved without any distribution-matching loss in training.
\begin{figure}[htbp!]
\centering
\includegraphics[width=\textwidth]{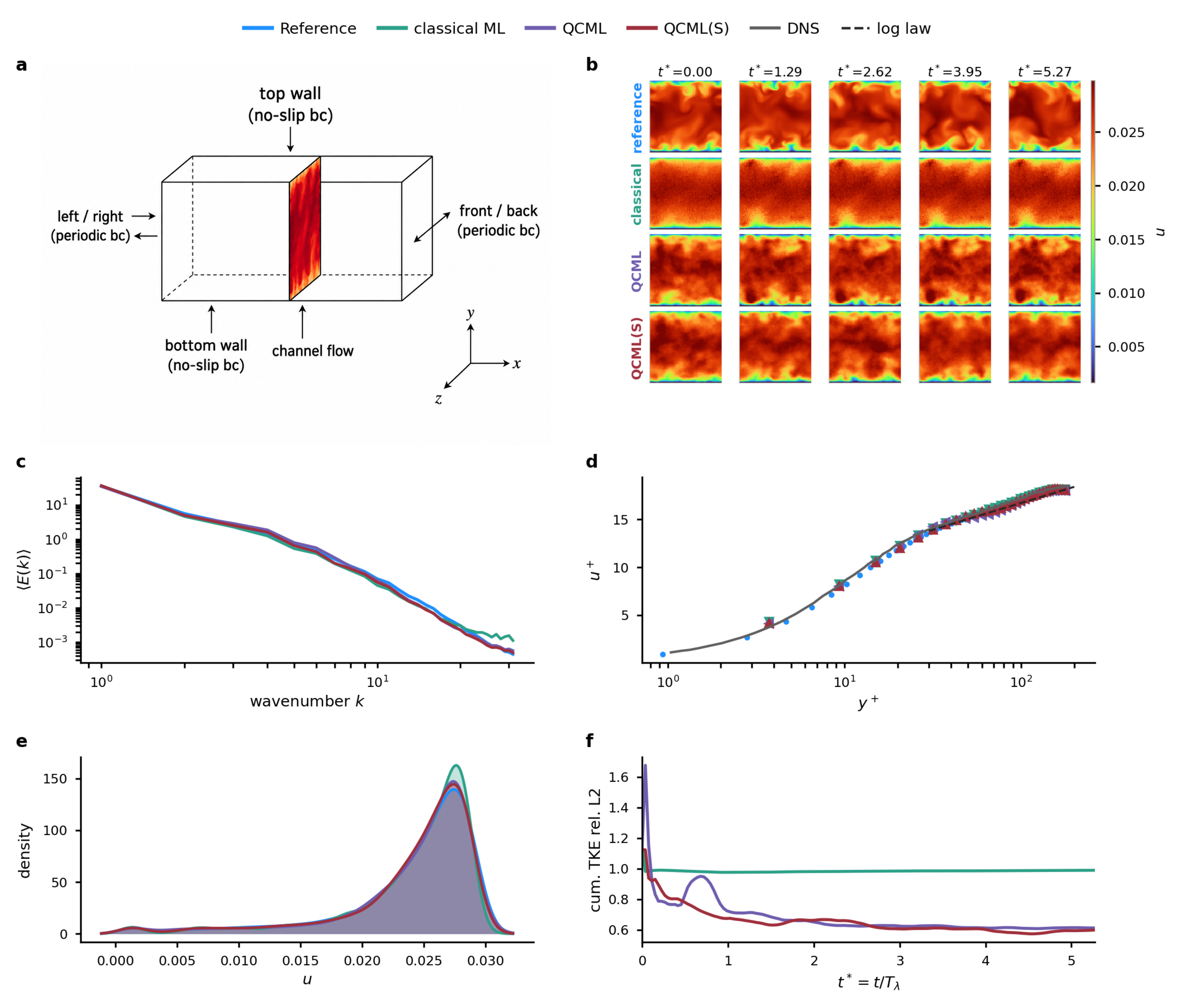}
\caption{\textbf{QCML on long-rollout turbulent channel flow.}
Blue, green, purple and red denote the reference data, classical ML, QCML and QCML (S), respectively; in panel \textbf{d}, grey and black dashed curves indicate DNS and the logarithmic law.
(\textbf{a}) Three-dimensional channel setup: no-slip top and bottom walls, periodic streamwise/spanwise boundaries; the highlighted plane is the mid-plane on which the surrogate is learned.
(\textbf{b}) Instantaneous velocity-field snapshots at Lyapunov-normalised times $t^{\ast}\in\{0.00,\,1.29,\,2.62,\,3.95,\,5.27\}$, with rows showing, from top to bottom, the reference, classical ML, QCML and QCML (S).
(\textbf{c}) Time-averaged isotropic energy spectrum $\langle E(k)\rangle$ versus wavenumber $k$.
(\textbf{d}) Mean streamwise velocity profile $u^{+}$ versus wall-normal coordinate $y^{+}$, with DNS and the log law at $Re_\tau=180$ as auxiliary references.
(\textbf{e}) Pixel-wise velocity density on the mid-plane (LBU).
(\textbf{f}) Cumulative turbulent kinetic-energy error relative to the reference (relative $L_2$) versus $t^{\ast}$.}
\label{fig:turbulent}
\end{figure}

Figure~\ref{fig:turbulent}f tracks how the surrogates evolve through the rollout via the cumulative turbulent-kinetic-energy (TKE) error relative to the reference (definition in Supplementary Information~S6.2). Both QCML variants start above unity at $t^{\ast}=0$, decrease rapidly within the first Lyapunov time, apart from a brief transient excursion of QCML near $t^{\ast}\approx 0.7$, and settle to a plateau of ${\sim}0.6$ for the remainder of the rollout, indicating statistical convergence of the surrogate state toward the reference attractor. Classical ML, by contrast, stays pinned at unity throughout: a flat TKE error at $1$ corresponds to a vanishing predicted TKE, that is, a near-zero coherent-fluctuation field, consistent with the stagnant, mean-like snapshots already visible in Fig.~\ref{fig:turbulent}b. Notably, the classical baseline is itself trained with a soft unitarity penalty $\|\mathbf{K}^{\top}\mathbf{K}-\mathbf{I}\|_F^{2}$ on its latent propagator~\cite{wangxue2026quantum}, so its collapse is not an artefact of an unregularised comparison: a soft penalty is minimised only up to optimisation tolerance and leaves a residual spectral drift that compounds over the rollout, whereas the unitarity of $U_q$ holds exactly for every parameter value. The two trajectories therefore separate by mechanism rather than by tuning, and we show in Methods, Sec.~``Stability against exploding and vanishing gradients'' why QCML is norm-preserving by construction whereas a classical latent operator, even a spectrally regularised one, can collapse under autoregressive rollout.

\section{Stenotic aortic flow}\label{sec:stenosis}

Stenosis of the aorta concentrates the haemodynamic challenge onto a single focal lesion. A $40\%$ diameter reduction in the thoracic aorta accelerates the local jet to ${\sim}1.8\,\mathrm{m\,s^{-1}}$, destabilises the distal shear layer and produces a pressure drop across the throat~\cite{feiger2020accelerating,xue2025uncertainty}. This throat pressure drop is the primary clinical biomarker for haemodynamic severity, so a surrogate must be validated on three quantities at once: the global pressure morphology across the cardiac phase, the multi-scale energy cascade inherited from the disturbed shear layer and the localised pressure drop across the throat.

\begin{figure}[htbp!]
\centering
\includegraphics[width=0.85\textwidth]{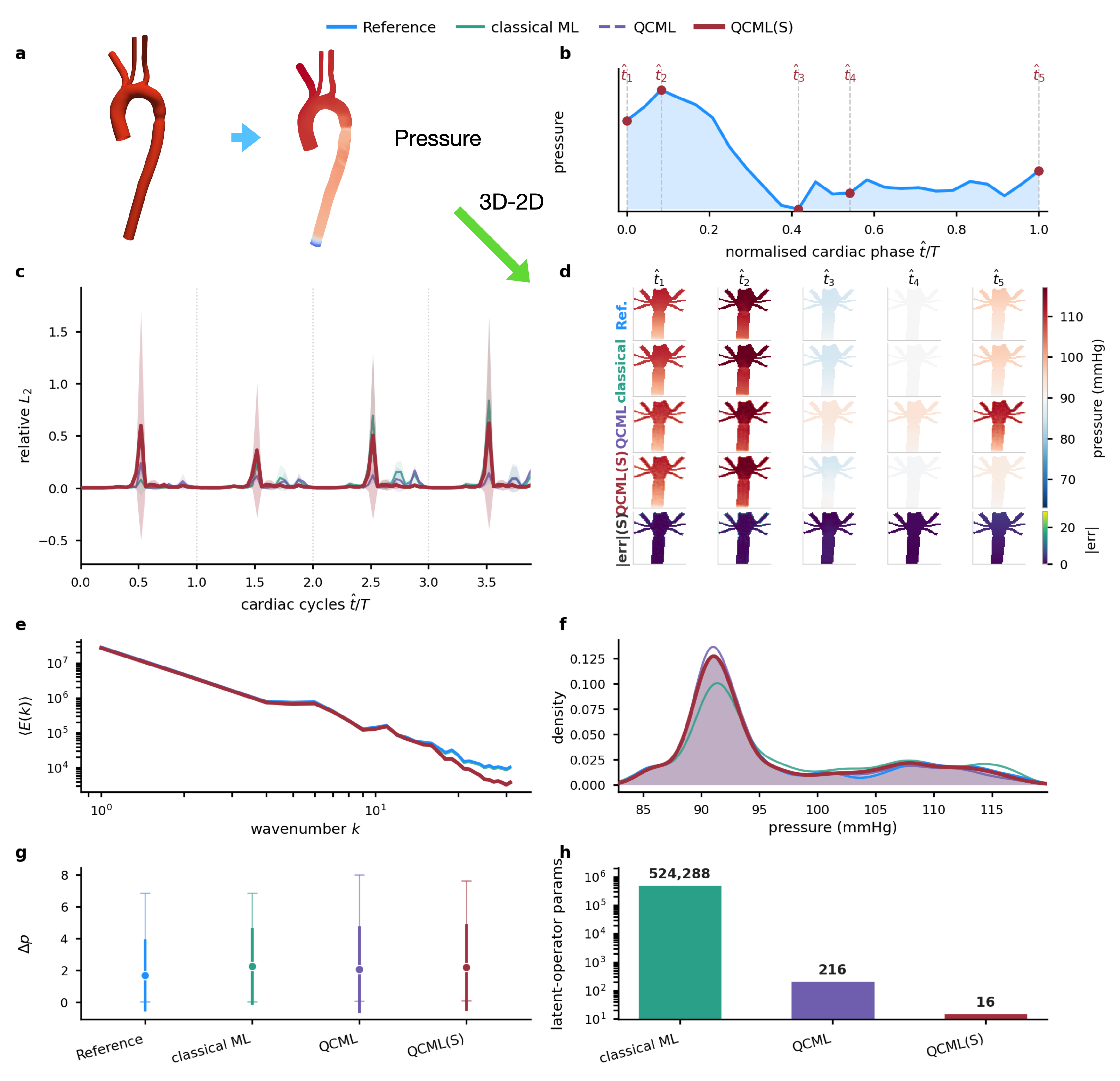}
\caption{\textbf{QCML on stenotic aortic haemodynamics.}
Line colours denote the reference, classical ML baseline, QCML and QCML (S) in blue, green, purple and red, respectively, unless otherwise stated.
(\textbf{a}) Patient-specific stenosis of the aorta geometry; the 3D pressure field is UV-unwrapped onto a 2D map across the throat ROI.
(\textbf{b}) Reference pressure waveform versus normalised cardiac phase $\hat{t}/T$, with five diagnostic instants $\hat{t}_1$--$\hat{t}_5$.
(\textbf{c}) Relative $L_2$ rollout error versus $\hat{t}/T$ over a ${\sim}3.9$-cycle horizon, normalised by $120\,\mathrm{mmHg}$; shaded bands are one s.d.\ across $N{=}3$ held-out trajectories, dotted vertical lines mark integer cycles.
(\textbf{d}) Per-phase pressure fields at $\hat{t}_1$--$\hat{t}_5$ (columns), with rows showing the reference, classical ML, QCML, QCML (S), and the absolute-error map $|\mathrm{err}|(\mathrm{S})$ of QCML (S) (mmHg).
(\textbf{e}) Time-averaged radial energy spectrum $\langle E(k)\rangle$ versus wavenumber $k$.
(\textbf{f}) Pixel-wise pressure density over all valid pixels and rollout frames (mmHg).
(\textbf{g}) Throat-ROI pressure drop $\Delta p$ pooled over $N{=}3$ trajectories and ${\sim}3.9$ cycles, for the reference, classical ML and both QCML variants; markers denote the mean, thick whiskers $\pm 1$ s.d., thin whiskers the full range.
(\textbf{h}) Trainable latent-propagator parameter count for classical ML, QCML and QCML (S).}
\label{fig:stenosis}
\end{figure}

Figure~\ref{fig:stenosis}a sketches the simulation and surrogate-modelling pipeline~\cite{xue2024lattice}. Pulsatile blood flow through a patient-specific thoracic aorta~\cite{Wilson2013TheResults} with a localised $40\%$ stenotic throat was simulated with the HemeLB lattice Boltzmann solver~\cite{mazzeo2008hemelb,zacharoudiou2023development,succi2001lattice} at $100\,\mu\mathrm{m}$ resolution, producing twenty independent trajectories of five cardiac cycles each, i.e.\ $100$ cardiac cycles of haemodynamic data in total. The three-dimensional surface pressure field on the curved aortic wall was then UV-unwrapped onto a two-dimensional map across the throat region of interest (ROI), and this 2D map is the field that all surrogates predict. The twenty trajectories were partitioned $75/10/15$ into independent training, validation and test sets; the full data-generation protocol, including the $20\,\mu\mathrm{m}$ DNS reference used to validate the LES-LBM resolution, is given in Supplementary~S2.2. Five representative cardiac instants $\hat{t}_1$--$\hat{t}_5$, spanning systolic acceleration, peak systole, distal shear-layer formation, late systole and diastolic relaxation, are marked along the pulsatile pressure waveform in Fig.~\ref{fig:stenosis}b. All surrogates were evaluated under the same autoregressive window-rollout protocol $\hat{\mathbf{u}}_{t+1:t+k} = f_{\theta}(\mathbf{u}_{t-k+1:t})$ as in Sec.~\ref{sec:turbulent} (Supplementary~S3.5), advanced from a reference initial window over ${\sim}3.9$ cardiac cycles without further reference input.

Figure~\ref{fig:stenosis}c shows that both QCML variants maintain low relative $L_2$ rollout error (Supplementary~S6.1) over multi-cycle prediction, on a par with the classical ML baseline. Away from brief phase-transition intervals, the error remains at the few-percent level, and the occasional sharp excursions are temporally localised and rapidly decay within the same cardiac cycle. The QCML and QCML (S) curves exhibit nearly overlapping envelopes throughout the rollout, indicating that the structured ansatz preserves the long-horizon accuracy of QCML without introducing monotonic error accumulation. Figure~\ref{fig:stenosis}d provides a point-wise comparison at five representative cardiac phases $\hat{t}_1$--$\hat{t}_5$. Classical ML and both QCML variants reproduce the systolic, diastolic and late-cycle pressure morphology of the reference, and the absolute-error map of the structured variant, $|\mathrm{err}|(\mathrm{S})$, remains much smaller than the local pressure amplitude, with residuals mainly confined to the stenotic jet region.

Spectral and statistical diagnostics over the same rollout corroborate this agreement. Figure~\ref{fig:stenosis}e shows that the time-averaged spectrum $\langle E(k)\rangle$ of both QCML variants tracked the reference over more than a decade of wavenumber space, with the two variants indistinguishable across the entire $k$-range. Figure~\ref{fig:stenosis}f shows that the pressure density of QCML (S) overlapped the reference within line width across the whole pressure range, including both the dominant peak near $90\,\mathrm{mmHg}$ and the high-pressure shoulder above $105\,\mathrm{mmHg}$ associated with peak systolic loading. QCML tracked the same distribution with only a slight overshoot at the main peak, whereas the classical ML baseline underestimated the peak density by around a fifth, showing that both QCML variants capture the pressure statistics more faithfully than the classical baseline, without any distribution-matching loss during training. Figure~\ref{fig:stenosis}g then evaluates the clinically decisive endpoint, the throat pressure drop $\Delta p = \bar{p}_{\mathrm{upstream}} - \bar{p}_{\mathrm{downstream}}$, defined as the cross-sectionally averaged pressure on the upstream side of the throat minus that on the downstream side (Supplementary~S6.6). The pooled $\Delta p$ distributions of both QCML variants reproduce the reference mean and interquartile range; QCML (S) sits marginally closer to the reference median and spread than QCML, although the difference is small relative to sampling variability. 
The parameter budget then separates the two variants in Fig.~\ref{fig:stenosis}h: the quantum-circuit parameter count drops from $216$ in QCML to $16$ in QCML (S), a ${\sim}13.5\times$ compression of the latent propagator at predictive accuracy that is essentially indistinguishable across Figs.~\ref{fig:stenosis}c--\ref{fig:stenosis}g. The two variants use different ans\"atze (Supplementary~S3.4): the generic QCML circuit carries $3LN_q$ trainable parameters, a general single-qubit rotation on every qubit plus a fixed entangling ring per layer ($L=8$ layers and $N_q=9$ qubits give $216$), whereas the structured QCML (S) keeps only $2L=16$ scalars, one shared phase $\alpha_\ell$ and one shared coupling $\beta_\ell$ per layer; the full structured ansatz and its layer-by-layer gate-level definition are given in Supplementary~S4. The contrast isolates the contribution of the structured ansatz: imposing physical structure on the quantum latent removes more than an order of magnitude of trainable parameters without measurable cost on either point-wise fields or the haemodynamic biomarker.

\section{Abdominal aortic aneurysm haemodynamics}\label{sec:aaa}

The abdominal aortic aneurysm (AAA) is the most clinically demanding setting in our evaluation. The flow is strongly pulsatile, the geometry is patient-specific, and the quantities most relevant to rupture-risk assessment, peak systolic pressure and local wall shear stress (WSS), are spatially concentrated and temporally transient~\cite{lo2024uncertainty,tanade2022analysis}. A surrogate must therefore be accurate not only on average, but in the precise spatiotemporal regions where haemodynamic risk is highest.

\begin{figure}[p]
\centering
\includegraphics[width=0.85\textwidth,height=0.85\textheight,keepaspectratio]{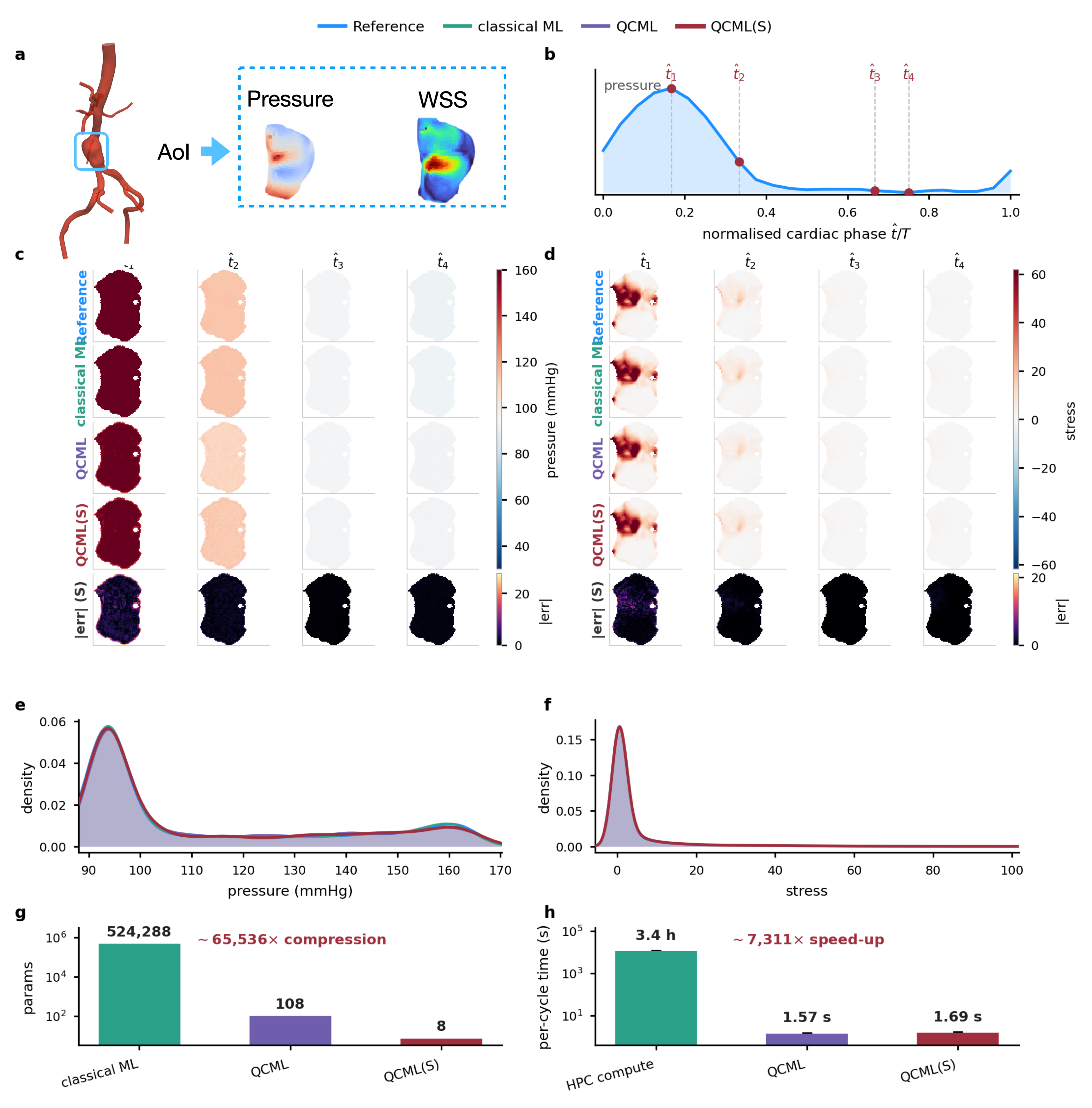}
\caption{\textbf{QCML on patient-specific abdominal aortic aneurysm haemodynamics.}
Line colours denote the reference, classical ML baseline, QCML and QCML (S) in blue, green, purple and red, respectively; field panels use the indicated pressure, stress and absolute-error colour bars.
(\textbf{a}) Patient-specific abdominal aortic aneurysm geometry; the 3D surface pressure and WSS fields on the area of interest (AoI) are UV-unwrapped onto matching 2D maps.
(\textbf{b}) Reference pressure waveform versus normalised cardiac phase $\hat{t}/T$, with four diagnostic instants $\hat{t}_1$--$\hat{t}_4$.
(\textbf{c}) Pressure fields over the AoI at $\hat{t}_1$--$\hat{t}_4$ (columns), with rows showing the reference, classical ML, QCML, QCML (S), and the absolute-error map $|\mathrm{err}|(\mathrm{S})$ of QCML (S) (mmHg).
(\textbf{d}) WSS fields at the same instants: same row layout as (c).
(\textbf{e}) Pixel-wise pressure density over the AoI (mmHg).
(\textbf{f}) Pixel-wise WSS-magnitude density over the AoI.
(\textbf{g}) Trainable latent-propagator parameter count for classical ML, QCML and QCML (S); the annotated ratio is the compression of QCML (S) relative to classical ML.
(\textbf{h}) Per-cycle latent-propagation cost for HPC compute, QCML and QCML (S), evaluated on a classical emulator of the quantum circuit.}
\label{fig:aaa}
\end{figure}

Figure~\ref{fig:aaa}a sketches the simulation and surrogate-modelling pipeline. Pulsatile blood flow through a single patient-specific AAA geometry~\cite{lo2024uncertainty}, with one main aortic inlet, eight side-branch inlets and two outlets, was simulated with the HemeLB lattice Boltzmann solver~\cite{mazzeo2008hemelb,zacharoudiou2023development} at $100\,\mu\mathrm{m}$ resolution, producing twenty independent trajectories generated by uniformly shifting the inlet-waveform phase across the diastolic low-velocity window. Each trajectory advances for five cardiac cycles, yielding $100$ cardiac cycles of haemodynamic data in total. The aneurysm area of interest (AoI) was extracted from the full 3D geometry, and the 3D surface pressure and WSS fields on the AoI were UV-unwrapped onto matching 2D maps, which are the fields all surrogates predict. The twenty trajectories were partitioned $80/10/10$ into independent training, validation and test sets; the full data-generation protocol, including the branch boundary conditions and the sponge-layer configuration used to damp outflow acoustic reflections~\cite{guo1994comparison,adams2000direct,lo2025multi}, is given in Supplementary~S2.3. Four representative cardiac instants $\hat{t}_1$--$\hat{t}_4$, spanning systolic acceleration, peak systole, late systole and diastolic relaxation, are marked along the pulsatile inflow waveform in Fig.~\ref{fig:aaa}b. QCML, QCML (S) and the classical ML baseline were evaluated under the same autoregressive rollout protocol $\hat{\mathbf{u}}_{t+1:t+k} = f_{\theta}(\mathbf{u}_{t-k+1:t})$ (Supplementary~S3.5), with the predicted field $\mathbf{u}_t = (p_t,\,\tau^{\mathrm{w}}_t)$ jointly comprising the AoI pressure and WSS, and advanced from a reference initial window over five cardiac cycles without further reference input.

Figure~\ref{fig:aaa}c compares the predicted pressure fields at $\hat{t}_1$--$\hat{t}_4$ against the reference. Both QCML and QCML (S) reproduced the elevated systolic field at $\hat{t}_1$ and the low diastolic field at $\hat{t}_3$--$\hat{t}_4$ in magnitude and spatial localisation, and the absolute-error map $|\mathrm{err}|(\mathrm{S})$ of QCML (S) stayed an order of magnitude below the local field amplitude. Figure~\ref{fig:aaa}d shows the WSS fields at the same instants: the peak stress concentration on the aneurysm wall at $\hat{t}_1$, which dominates rupture-risk evaluation, was recovered by both QCML variants with the correct hotspot location and intensity, with residuals concentrated outside the clinically critical region.

Distribution-level diagnostics over the same rollout corroborate this agreement; the formal definitions of the AAA biomarkers are given in Supplementary Information~S6.5 (pixel-wise PDFs) and S6.7 (peak pressure and peak wall shear stress). Figure~\ref{fig:aaa}e shows that the pressure density aggregated over the AoI traces the bimodal reference distribution for both QCML variants, including the high-pressure tail above $160\,\mathrm{mmHg}$ associated with peak systolic loading; the two prediction curves overlap the reference within line width. Figure~\ref{fig:aaa}f shows that both QCML variants reproduced the heavy-tailed WSS density of the reference, dominated by a sharp peak near baseline stress with a long tail toward elevated values, again with no distribution-matching loss during training.

The combination of parameter efficiency and inference speed is the most striking result. Figure~\ref{fig:aaa}g shows that the classical baseline carries $524{,}288$ trainable latent parameters, QCML uses $108$ and QCML (S) only $8$, a ${\sim}65{,}536\times$ compression of the latent propagator from classical ML to the structured QCML variant at predictive accuracy that is indistinguishable across Figs.~\ref{fig:aaa}c--\ref{fig:aaa}f. The three models use different latent propagators (Supplementary~S3.4): classical ML uses a dense Koopman operator on a $D=512$ latent space, trained with a soft unitarity regulariser ($2D^2\approx 5.24\times 10^{5}$ entries across the forward and backward operators), the generic QCML circuit carries $3LN_q$ trainable parameters ($L=4$ layers and $N_q=9$ qubits give $108$), and the structured QCML (S) keeps only $2L=8$ scalars, one shared phase and one shared coupling per layer; the full structured ansatz and its gate-level definition are given in Supplementary~S4. Figure~\ref{fig:aaa}h reports the per-cycle latent-propagation cost, evaluated on a classical emulator of the quantum circuit. Relative to the ${\sim}3.4\,\mathrm{h}$ required per cardiac cycle by the HPC CFD reference on $1{,}024$ CPU cores, both QCML variants reduce this algorithmic cost by several orders of magnitude. This figure quantifies the intrinsic inference complexity of the surrogate and deliberately excludes the state-preparation, measurement-shot and classical--quantum data-transfer overheads that would dominate an end-to-end run on present quantum hardware (Supplementary~S3.6); it is therefore not a hardware wall-clock claim. Figure~\ref{fig:aaa}g, the parameter compression, is a hardware-independent property of the model and carries the primary efficiency message: imposing physical structure on the quantum latent removes more than four orders of magnitude of trainable parameters without measurable cost on either point-wise fields or the haemodynamic distributions.

\section{Noise-aware emulator and hardware validation}\label{sec:hardware_validation}

The results in Figs.~\ref{fig:turbulent}--\ref{fig:aaa} were obtained with a classical emulator of the structured quantum circuit. To assess backend robustness, the trained QCML (S) models were evaluated either on IQM's 54-qubit Emerald processor or on a noise-aware emulator configured with the same device-noise profile (See Methods, Sec.~``Hardware implementation and training''). Table~\ref{tab:hardware_validation} reports the backend and the one-step prediction agreement for each benchmark. Executed on Emerald, the turbulent-channel model retained $79.69\%$ agreement with the reference under device noise. The stenotic-aorta and abdominal-aortic-aneurysm models, evaluated on the noise-aware emulator inside the same inference loop, reached $95.54\%$ and $95.25\%$ agreement respectively. Backend noise therefore degrades the one-step prediction but does not destroy it, and the trained models transfer across backends without retraining.

\begin{table}[htbp!]
\centering
\caption{Backend validation of QCML (S) on the three benchmark flows. Agreement is defined as the complement of the one-step relative $L_2$ prediction error, $\mathrm{agreement}=(1-\ell_{\mathrm{rel}\,L_2})\times 100\%$, where $\ell_{\mathrm{rel}\,L_2}=\|\hat{y}-y\|_2/\|y\|_2$ compares the predicted next frame with the reference next frame. Losses are summed within each batch and then averaged over the test-loader batches of the held-out trajectories; identity, backward and long-rollout losses are not included. The turbulent-channel case was evaluated on IQM's 54-qubit Emerald superconducting quantum processor; the stenotic-aorta and abdominal-aortic-aneurysm cases were evaluated on a noise-aware emulator configured with the Emerald device-noise profile.}
\label{tab:hardware_validation}
\begin{tabular}{l l @{\hspace{3em}} l}
\toprule
Benchmark (diagnostic)                                    & Backend        & Agreement       \\
\midrule

Turbulent channel flow inflow          & Emerald QPU    & $79.69\%$       \\
Stenotic aorta                               & Noise-aware emulator & $95.54\%$ \\
Abdominal aortic aneurysm    & Noise-aware emulator & $95.25\%$ \\
\bottomrule
\end{tabular}
\end{table}

\section{Conclusion}\label{sec:conclusion}

We have introduced quantum-compressed machine learning as a hybrid quantum-classical surrogate for nonlinear physical flows. A classical encoder-decoder learns a compact representation of the physical state, and a structured quantum propagator carries that representation forward in time on a small superconducting quantum processor. We evaluated the framework on three benchmarks of increasing complexity: turbulent channel flow under extended autoregressive rollouts, stenotic aortic flow with its associated pressure drop, and patient-specific abdominal aortic aneurysm haemodynamics. Across these benchmarks, QCML does more than reproduce the classical baseline: it maintains coherent turbulent structures over five Lyapunov times after the soft-unitarity-regularised classical model collapses, captures stenotic pressure statistics more faithfully, and preserves pressure and wall-shear-stress fidelity in patient-specific aneurysm haemodynamics, while reducing the latent-propagator parameter count from $524{,}288$ to ${\sim}108$ in QCML and to as few as $8$ in QCML (S). On a classical emulator of the quantum circuit, per-cycle inference takes ${\sim}1.69\,\mathrm{s}$ for QCML (S), compared with ${\sim}3.4\,\mathrm{h}$ per cardiac cycle for conventional CFD on $1{,}024$ CPU cores, corresponding to a ${\sim}7{,}300\times$ reduction in algorithmic inference cost. 

Three advantages underlie these results. Long-horizon stability follows from the unitarity of the latent propagator, which pins its spectrum to the unit circle by construction and suppresses the compounding term that dominates autoregressive surrogate error; the turbulent-channel benchmark shows that approximate spectral regularisation does not substitute for this exact constraint. Parameter efficiency follows from the structured circuit, whose mode-wise rotations and sparse inter-mode couplings compress the latent propagator by more than four orders of magnitude without measurable loss of expressive power, as the structured variant matches the unshared QCML on the haemodynamic biomarkers. Explainability follows from the same structure: the trainable parameters map one-to-one onto identifiable mode frequencies and inter-mode interactions, in contrast to the opaque dense matrices of neural latents.

Three directions remain open for scaling the framework towards deployment: tighter integration of quantum processors with HPC clusters over low-latency interconnects, to reduce the classical–quantum data-exchange overhead, which becomes limiting at larger circuit sizes; structured circuits with higher qubit counts and richer connectivity graphs, opening a path from the two-dimensional projections treated here to volumetric, patient-specific flows; and the transition to early fault-tolerant processors, which will lift the noise floor that currently limits hardware fine-tuning depth. For safety-relevant nonlinear systems, where transparent latent dynamics are a deployment prerequisite rather than an afterthought, a surrogate that is at once parameter-efficient, structurally stable and directly explainable offers a credible path towards trustworthy hybrid quantum-classical prediction.

\section*{Methods}

\subsection{Hybrid quantum--classical architecture}

QCML reformulates the Koopman learning paradigm by realising the latent evolution of dynamical systems within a quantum-mechanical representation. Instead of representing the learned linear dynamics as an explicit classical matrix operator, it implements latent propagation through a structured unitary transformation acting in a quantum Hilbert space, followed by observable-based reconstruction in the latent space.

For a discrete-time dynamical system
\begin{equation}
\mathbf{u}_{t+1} = f(\mathbf{u}_t),
\end{equation}
the Koopman operator $\mathcal{K}$ acts linearly on observables $g:\mathcal{M}\rightarrow\mathbb{C}$ as
\begin{equation}
(\mathcal{K}g)(\mathbf{u}) = g(f(\mathbf{u})),
\end{equation}
where $\mathcal{M}\subset\mathbb{R}^d$ denotes the physical state space. In data-driven Koopman learning, an encoder-decoder pair $(\phi,\psi)$ constructs a finite-dimensional latent representation in which nonlinear dynamics are approximately linear:
\begin{equation}
\mathbf{z}_{t+1} \approx U_K \mathbf{z}_t, 
\qquad 
\hat{\mathbf{u}}_t = \psi(\mathbf{z}_t),
\label{eq:koopman_linear}
\end{equation}
where $\mathbf{z}_t = \phi(\mathbf{u}_t) \in \mathcal{Z}$ is the latent coordinate and $U_K$ approximates the Koopman propagator in the latent space $\mathcal{Z}$.

In QCML, the latent space $\mathcal{Z}$ is embedded into a quantum Hilbert space $\mathcal{H}_{N_q}$ consisting of $N_q$ qubits. Under amplitude encoding, the latent dimension satisfies $k = 2^{N_q}$~\cite{gonzalez2024efficient}. The latent state $\mathbf{z}_t$ is mapped to a quantum state $|\psi_{\mathrm{in}}(\mathbf{z}_t)\rangle \in \mathcal{H}_{N_q}$, and its evolution is implemented through a unitary operator $U_q(\boldsymbol{\theta})$ acting on the quantum register:
\begin{equation}
|\psi_{\mathrm{out}}\rangle 
= U_q(\boldsymbol{\theta}) |\psi_{\mathrm{in}}(\mathbf{z}_t)\rangle,
\qquad
\mathbf{z}_{t+1}
= \mathcal{R}\!\left(|\psi_{\mathrm{out}}\rangle\right),
\label{eq:quantum_koopman}
\end{equation}
where $\mathcal{R}(\cdot)$ denotes the observable-based reconstruction map, implemented through expectation values of a fixed set of measurement observables $\hat{\mathbf{O}}$ (see Supplementary Information~S4 for the explicit choice).

The headline propagator, QCML (S), is constructed from a structured circuit. Rather than employing a generic hardware-efficient ansatz, we restrict the generators of the evolution to operators corresponding to mode-wise single-qubit rotations and structured inter-mode coupling. Specifically, the structured unitary operator is drawn from the family
\begin{equation}
U_q(\boldsymbol{\theta}) =
\prod_{\ell=1}^{L}
\left(
\prod_{i=1}^{N_q} e^{-i\alpha_{\ell,i} X_i}
\right)
\left(
\prod_{(i,j)\in E} e^{-i\beta_{\ell,(i,j)}(X_iX_j+Y_iY_j)}
\right),
\label{eq:Uq_circuit}
\end{equation}
where $X_i$ generates a mode-wise phase rotation of the $i$-th latent mode and $X_iX_j + Y_iY_j$ induces structured coupling between modes adjacent on a sparse graph $E$. QCML (S) ties these parameters within each layer, sharing one phase $\alpha_{\ell}$ across all qubits and one strength $\beta_{\ell}$ across all edges, so the trainable set reduces to $\boldsymbol{\theta}=\{\alpha_{\ell},\beta_{\ell}\}_{\ell=1}^{L}$ with only $2L$ scalars (Fig.~\ref{fig:qcml_device}b). The baseline variant QCML instead uses a generic hardware-efficient circuit, a general single-qubit rotation (three Euler angles) on every qubit followed by a fixed entangling ring in each layer, giving $3LN_q$ trainable parameters and no trainable couplings; comparing the two isolates the contribution of the imposed structure.

The unitarity of the quantum propagator is guaranteed by construction, $U_q(\boldsymbol{\theta})^{\dagger} U_q(\boldsymbol{\theta}) = I$, which preserves the norm of the encoded quantum state:
\begin{equation}
\langle \psi_{\mathrm{out}} | \psi_{\mathrm{out}} \rangle
=
\langle \psi_{\mathrm{in}}(\mathbf{z}_t) | \psi_{\mathrm{in}}(\mathbf{z}_t) \rangle .
\end{equation}
All eigenvalues of $U_q(\boldsymbol{\theta})$ therefore lie on the unit circle, $\lambda_j=e^{i\omega_j}$ for real phases $\omega_j$, so the quantum latent state undergoes a norm-preserving evolution in Hilbert space. The centre inset of Fig.~\ref{fig:framework}a illustrates this behaviour through a single-qubit Bloch-sphere schematic, whose equatorial $(x,y)$ plane encodes the real and imaginary parts of the off-diagonal coherence. This unitary backbone replaces the exponential rollout error of generic linear latent propagators with a strictly linear-in-$n$ bound, suppressing in principle the unstable amplification and decay regimes that destabilise autoregressive surrogates over long horizons.

\subsection{Stability of the quantum latent propagator in autoregressive rollout}

The deployed surrogate produces predictions by composing encoder, latent propagator and decoder, and is then advanced autoregressively on its own outputs, so the same latent propagator $U$ is applied $n$ times across an $n$-step rollout. For any observable $g$ in the encoder's $N$-dimensional subspace $\mathcal{V}_N\subset\mathcal{H}$, the gap between the true Koopman evolution $\mathcal{K}^n g$ and the surrogate composition $U^n g$ obeys the Koopman--Galerkin telescoping bound (Supplementary Information~S5.2)
\begin{equation}
\bigl\|\mathcal{K}^{n} g - U^{n} g\bigr\|_{\mathcal{H}}
\;\le\;
\varepsilon_N \sum_{j=0}^{n-1} \rho(U)^{\,j},
\label{eq:rollout_telescope}
\end{equation}
where $\varepsilon_N$ is the per-step encoder--decoder invariance defect~\cite{williams2015data,korda2018convergence} and $\rho(U)$ is the spectral radius of the latent propagator. The full encoder--latent--decoder error therefore decomposes into a per-step \emph{additive} encoder--decoder reconstruction term $\varepsilon_N$ and a \emph{multiplicative} latent compounding $\rho(U)^n$; long-horizon stability is controlled by the latter, because $\varepsilon_N$ is fixed per step while $\rho(U)^n$ accumulates geometrically with $n$. The autoregressive question therefore reduces to a single question: how fast does $\|U^{n}\|_2$ grow with $n$?

This compounding of a single operator $U$ over $n$ rollout steps is structurally analogous to the mechanism behind exploding and vanishing gradients in autoregressive sequence models~\cite{bengio1994,pascanu2013}: the forward latent state and the gradient back-propagated from the loss $\mathcal{L}$ both pass through powers of the same operator,
\begin{equation}
\mathbf{z}_n = U^{n}\mathbf{z}_0,
\qquad
\frac{\partial \mathcal{L}}{\partial \mathbf{z}_0}
= \bigl(U^{\dagger}\bigr)^{n}\frac{\partial \mathcal{L}}{\partial \mathbf{z}_n}.
\label{eq:bptt}
\end{equation}
The analogy is structural only: our model is not a recurrent network, since the latent state is a Koopman observable obtained by the encoder rather than a learned hidden vector, and training does not back-propagate through a learned recurrent dynamics. What carries over is the mathematical issue of repeated $U$-composition. For a general, possibly non-normal $U$, the growth of $\|U^{n}\|_2$ cannot be read off any single singular value at finite $n$, and intermediate steps may exhibit transient amplification even when $U$ is ultimately contractive. The asymptotic rate is nevertheless pinned by a classical theorem of Gelfand~\cite{gelfand1941}, which identifies it with the spectral radius,
\begin{equation}
\rho(U)\;=\;\lim_{n\to\infty}\|U^{n}\|_2^{1/n},
\label{eq:gelfand}
\end{equation}
so the long-horizon rollout norm falls into a sharp trichotomy,
\begin{equation}
\|U^{n}\mathbf{z}_0\|_2
\;\underset{n\to\infty}{\sim}\;
\rho(U)^{\,n}\,\|\mathbf{z}_0\|_2,
\qquad
\rho(U) \;=\;
\begin{cases}
1+\delta>1, & \text{exploding state and gradient},\\[2pt]
1,          & \text{norm-preserving},\\[2pt]
1-\delta<1, & \text{vanishing state and gradient}.
\end{cases}
\label{eq:rollout_dichotomy}
\end{equation}
The asymptotic equivalence is up to polynomial factors arising from any non-trivial Jordan structure of $U$; equality on the middle row requires $U$ to be normal, which is automatic for the unitary $U_q$ used below. For any measure-preserving dynamical system, the Koopman operator is an $L^{2}$ isometry with $\rho(\mathcal{K})=1$. A surrogate with $\rho<1$ then injects spurious dissipation into the encoder--latent--decoder loop, and one with $\rho>1$ injects spurious amplification. The physically faithful target is therefore not merely a contractive $U$, but one whose spectrum sits on the unit circle.

\paragraph{Classical latent propagators.}
A classical dense propagator $U_K$ is typically fitted by least squares on single-step prediction, which controls only the one-step residual and imposes no global constraint on the eigenvalue distribution of $U_K$. Spectral discipline can be encouraged by augmenting the training loss with a soft unitarity penalty $\|U_K^{\top}U_K - I\|_F^{2}$, and the classical ML baseline used throughout this work carries exactly such a regulariser on its latent propagator~\cite{wangxue2026quantum}. A penalty of this form is, however, a regularisation target rather than a constraint: it competes with the data-fit terms and is minimised only up to the optimisation tolerance, so at convergence it is small but generically non-zero and the trained propagator retains a residual spectral drift $\delta>0$, leaving $\rho(U_K)$ detached from the unit circle. A residual of only $\delta = 10^{-2}$ already compounds to $(1+\delta)^{n}\approx e^{\delta n} \approx 148$ over $n = 500$ autoregressive inference steps. Both exploding and vanishing rollout regimes therefore remain generic failure modes of classical autoregressive surrogates, with or without soft spectral regularisation. Specialising Eq.~\eqref{eq:rollout_telescope} to $\rho = 1+\delta > 1$ and assuming $U_K$ is normal so that $\|U_K\|_2 = \rho$, the geometric sum is dominated by its highest power and the bound becomes
\begin{equation}
\bigl\|\mathcal{K}^{n} g - U_K^{n} g\bigr\|_{\mathcal{H}}
\;\le\;
\varepsilon_N\,\frac{\rho^{\,n}-1}{\rho-1}
\;\sim\;
\frac{\varepsilon_N}{\delta}\,e^{\delta n}
\qquad (\rho = 1+\delta > 1),
\label{eq:classical_error}
\end{equation}
where $\varepsilon_N$ is the encoder-decoder invariance defect. (For non-normal $U_K$, a numerical-range constant must be added; the bound is otherwise unchanged.)

\paragraph{Quantum latent propagator.}
Lyapunov's theory of stability~\cite{lyapunov1892general,khalil2002nonlinear} provides the canonical criterion for bounded trajectories of a discrete-time system $\mathbf{z}_{t+1}=\Phi(\mathbf{z}_t)$. It suffices to exhibit a continuous energy $V:\mathbb{C}^{k}\to\mathbb{R}_{\ge 0}$ with $V(\mathbf{0})=0$ and $V(\mathbf{z})>0$ for $\mathbf{z}\neq\mathbf{0}$ that does not increase along the flow, $V(\Phi(\mathbf{z}))\le V(\mathbf{z})$. When this inequality saturates as $V(\Phi(\mathbf{z}))= V(\mathbf{z})$, the system is marginally stable with conserved latent energy, and every eigenvalue of the one-step map lies on the unit circle. For the quantum latent propagator, this criterion is met by the simplest choice, the Euclidean latent energy $V(\mathbf{z}) = \|\mathbf{z}\|_2^{2}$, where $\mathbf{z}$ is identified with the amplitude vector of $|\psi_{\mathrm{in}}(\mathbf{z})\rangle$ under amplitude encoding. Because $U_q(\boldsymbol{\theta})$ is a product of exponentials of Hermitian generators (Eq.~\eqref{eq:Uq_hamiltonian} below) and is therefore unitary, a one-line computation gives
\begin{equation}
V\!\bigl(U_q(\boldsymbol{\theta})\,\mathbf{z}\bigr)
\;=\;
\mathbf{z}^{\dagger} U_q^{\dagger} U_q \,\mathbf{z}
\;=\;
\|\mathbf{z}\|_2^{2}
\;=\;
V(\mathbf{z})
\qquad \text{for every } \boldsymbol{\theta} \text{ and every } \mathbf{z},
\label{eq:lyapunov}
\end{equation}
so $V$ is a strict Lyapunov function with conserved energy for the latent rollout. Because Eq.~\eqref{eq:lyapunov} holds pointwise in parameter space, $\rho(U_q(\boldsymbol{\theta}))=1$ is not a target to be optimised towards but an identity imposed by the model class itself, holding at random initialisation, at every stochastic-gradient-descent (SGD) update, and at inference.

This Lyapunov--spectrum correspondence closes the chain of Eqs.~\eqref{eq:bptt}--\eqref{eq:rollout_dichotomy}. Conserved energy under $U_q$ pins every eigenvalue of $U_q$ to the unit circle. Setting $\rho(U_q)=1$ collapses the trichotomy of Eq.~\eqref{eq:rollout_dichotomy} onto its norm-preserving branch, ruling out both exploding and vanishing rollout regimes. Taking the limit $\rho\to 1$ in Eq.~\eqref{eq:classical_error} via $\lim_{\rho\to 1}(\rho^{\,n}-1)/(\rho-1) = n$ then replaces the exponential classical bound with a strictly linear one,
\begin{equation}
\bigl\|\mathcal{K}^{n} g - U_q^{n}(\boldsymbol{\theta})\, g\bigr\|_{\mathcal{H}}
\;\le\;
n\,\varepsilon_N
\qquad \text{for all } n \ge 1,
\label{eq:quantum_error}
\end{equation}
whose slope is the encoder-decoder invariance defect $\varepsilon_N$ alone. A self-contained proof of Eqs.~\eqref{eq:lyapunov}--\eqref{eq:quantum_error} is given in Supplementary Information~S5.2, together with the lemma that propagates the per-step defect $\varepsilon_N$ into the $n$-step bound and the corresponding corollary for classical dense propagators.

The layered circuit of Eq.~\eqref{eq:Uq_circuit} admits a Hamiltonian representation as a product of $L$ layer-wise unitaries,
\begin{equation}
U_q(\boldsymbol{\theta}) = \prod_{\ell=1}^{L} e^{-iH_{\ell}(\boldsymbol{\theta})\,\Delta t/L},
\label{eq:Uq_hamiltonian}
\end{equation}
in which each Hermitian generator $H_{\ell}(\boldsymbol{\theta}) = H_{\ell}(\boldsymbol{\theta})^{\dagger}$ governs the latent-space dynamics within a single layer,
\begin{equation}
H_{\ell}(\boldsymbol{\theta})
=
\sum_{i=1}^{N_q} \alpha_{\ell,i}\, X_i
+
\sum_{(i,j)\in E} \beta_{\ell,(i,j)}(X_iX_j + Y_iY_j),
\label{eq:layer_hamiltonian}
\end{equation}
combining latent-mode phase evolution with sparse inter-mode energy transfer, with QCML (S) tying $\alpha_{\ell,i}=\alpha_\ell$ and $\beta_{\ell,(i,j)}=\beta_\ell$ within each layer. The eigenvalues of each $H_{\ell}(\boldsymbol{\theta})$ define characteristic frequencies $\omega_j^{(\ell)}$ through $\lambda_j^{(\ell)} = e^{-i\omega_j^{(\ell)}\Delta t/L}$, yielding a layer-resolved spectral decomposition of the learned dynamics in Hilbert space. This structure aligns with the Koopman spectral viewpoint, in which nonlinear dynamics are approximated as a superposition of mode frequencies and structured mode interactions. Each layer-wise factor in Eq.~\eqref{eq:Uq_hamiltonian} is unitary, and so is their product $U_q(\boldsymbol{\theta})$; the Lyapunov argument of Eq.~\eqref{eq:lyapunov} therefore applies layer by layer and globally.

The structured quantum latent propagator of Eq.~\eqref{eq:Uq_hamiltonian}--\eqref{eq:layer_hamiltonian} delivers a decisive advantage over a classical dense Koopman propagator on the same latent. It is structurally immune to the exploding and vanishing rollout regimes that plague classical latent dynamics. Unitarity pins $\rho(U_q)=1$ pointwise in $\boldsymbol{\theta}$, replacing the exponential classical-latent error of Eq.~\eqref{eq:classical_error} with the strictly linear-in-$n$ Koopman--Galerkin bound of Eq.~\eqref{eq:quantum_error}. In parallel, amplitude encoding stores the $k$-dimensional latent in $N_q=\log_2 k$ qubits, so the dense matrix--vector multiplication
\begin{equation}
(U_K \mathbf{z}_t)_i
=
\sum_{j=1}^{k}
(U_K)_{ij} (\mathbf{z}_t)_j,
\end{equation}
of cost $\mathcal{O}(k^2)$ per step~\cite{li2020fourier} is replaced by the structured circuit of Eq.~\eqref{eq:Uq_circuit} at a tighter gate complexity
\begin{equation}
\mathrm{Cost}_{\mathrm{QCML}}
=
\mathcal{O}(L\,N_q)
=
\mathcal{O}(L\log k),
\label{eq:cost}
\end{equation}
linear in $\log k$ rather than polynomial in $k$.

The structured generator family and its Koopman spectral interpretation are detailed in Supplementary Information~S4. The Lyapunov stability of the latent rollout and the associated linear error bound are derived in Supplementary Information~S5.2, and the parametric-sensitivity and locality analyses follow in Supplementary Information~S5.3--S5.5.

\subsection{Hardware implementation and training}
\label{sec:hardware_implementation}

\begin{figure}[htbp]
\centering
\includegraphics[width=0.85\textwidth]{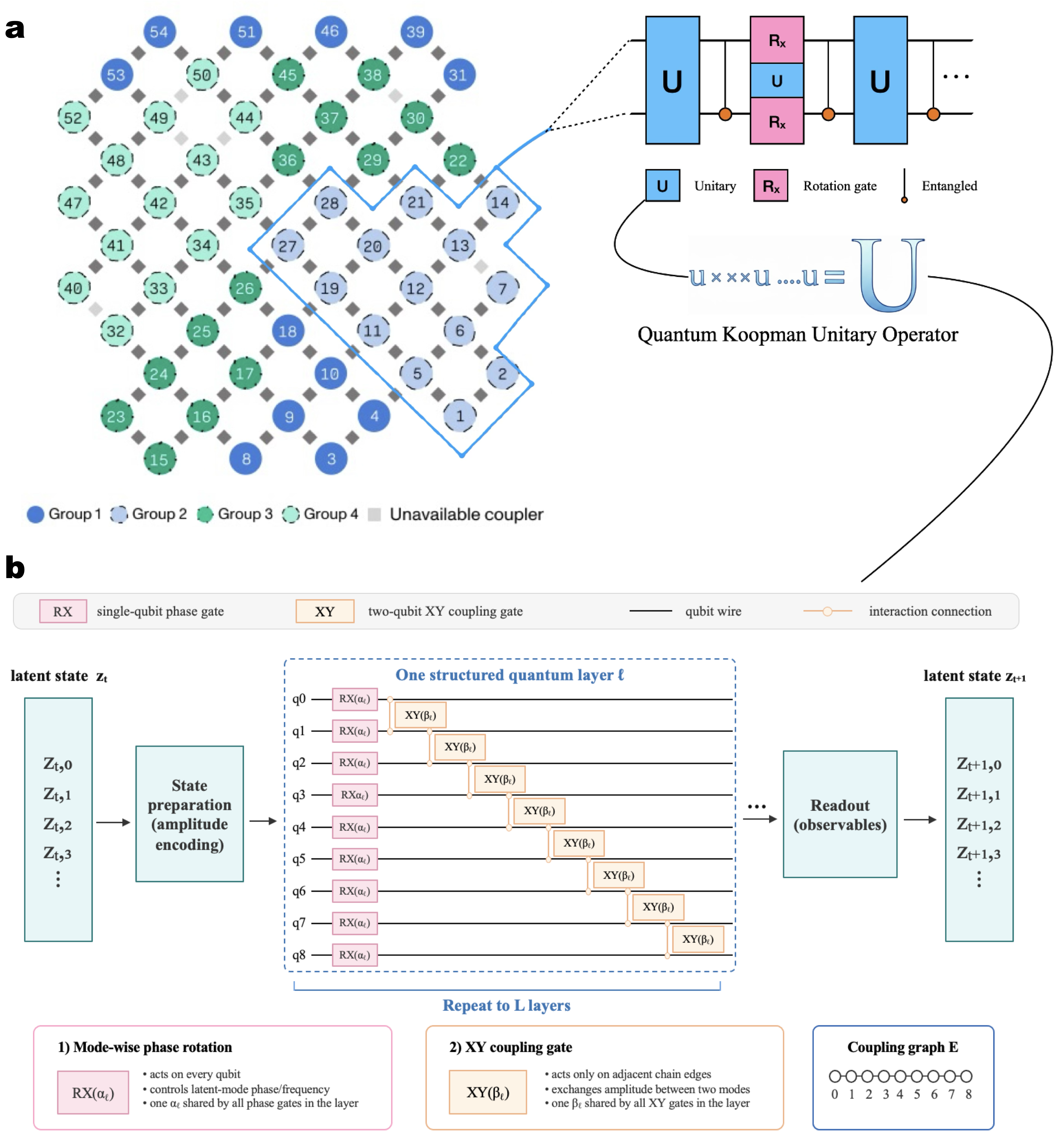}
\caption{\textbf{Structured quantum circuit and its hardware realisation.}
\textbf{a.} Native connectivity graph of the IQM 54-qubit Emerald processor; colours mark the four qubit groups, unavailable couplers are greyed out, and the blue outline highlights the qubit subset used in one execution of a single task. Each task was run as four parallel executions to reduce resource consumption.
\textbf{b.} Native gate-level realisation of the structured quantum latent propagator $U_q(\boldsymbol{\theta})$ of Eq.~\eqref{eq:Uq_circuit} after transpilation to the IQM Emerald native gate set: the latent state $\mathbf{z}_t$ is amplitude-encoded into the qubit register, transformed by $L$ repeated structured layers $U_\ell$, and read out into $\mathbf{z}_{t+1}$. Each layer uses the native $\mathrm{RX}(\alpha_\ell)$ single-qubit gate for the mode-wise rotation and the native $\mathrm{XY}(\beta_\ell)$ two-qubit gate for the XY coupling on adjacent qubits along the interaction graph $E$. In QCML (S), a single $\alpha_\ell$ and a single $\beta_\ell$ are shared across all $\mathrm{RX}$ and $\mathrm{XY}$ gates of layer~$\ell$, leaving $2L$ trainable scalars per circuit.}
\label{fig:qcml_device}
\end{figure}

The encoder, decoder and quantum-circuit parameters of Eqs.~\eqref{eq:koopman_linear}--\eqref{eq:quantum_koopman} are optimised jointly in a single end-to-end hybrid loop against a field-space reconstruction loss. Training is first carried out on a noise-aware classical emulator of the structured circuit to validate convergence. Backend validation is then performed either on the same noise-aware emulator or on IQM's superconducting quantum processor Emerald, a 54-qubit device for quantum latent propagator (Fig.~\ref{fig:qcml_device}).

Classical neural-network components are implemented in PyTorch and executed on the BEAST GPU cluster at the Leibniz Supercomputing Centre, while the quantum circuit is invoked through Qiskit interfaces to Emerald. The low gate count and shallow depth of the structured ansatz of Eq.~\eqref{eq:Uq_circuit} keep the hybrid execution within the coherence-limited regime of the device. The classical–quantum exchange between BEAST and IQM Emerald is carried over a dedicated network channel between the supercomputing centre and IQM, which keeps the round-trip overhead modest at the circuit sizes used here. Co-locating the QPU with the HPC system would reduce this overhead further.

\subsubsection{Structured quantum latent propagator}

The choice of variational quantum ansatz is central to the trainability of the quantum latent propagator. One of the main obstacles in variational quantum circuits is not back-propagation in the ordinary neural-network sense, but the emergence of barren plateaus in the cost landscape. As the number of qubits, the circuit depth or the expressibility of the ansatz increases, gradients of expectation-value objectives can concentrate around zero and their variance can decay exponentially~\cite{mcclean2018barren,cerezo2022challenges}. A key mechanism is concentration of measure: when an ansatz becomes sufficiently expressive to approximate Haar-random unitaries, or to form an approximate unitary 2-design, local parameter perturbations are averaged over the Hilbert space and have exponentially small influence on measured observables.

The design goal is therefore not simply to reduce the number of possible unitary maps. It is to avoid an unstructured search over the full unitary group while preserving the dynamical transformations needed by the problem. QCML (S) uses a structured ansatz for this purpose. The circuit restricts the latent propagator to the submanifold generated by mode-wise single-qubit rotations and sparse inter-mode couplings, which are the two operator classes required to represent latent frequencies and their leading interactions. This structure reduces exposure to one common route into barren plateaus, while keeping the search space aligned with the spectral dynamics that determine prediction accuracy.

The quantum latent propagator is implemented through the structured parameterised quantum circuit of Eq.~\eqref{eq:Uq_circuit}. Rather than an unstructured hardware-efficient ansatz of arbitrary rotations and entanglers, the framework restricts the generators to the mode-wise single-qubit rotations and sparse inter-mode couplings of Eq.~\eqref{eq:Uq_circuit}, with the interaction graph $E$ following the native connectivity of the quantum processor (Fig.~\ref{fig:qcml_device}a).

\subsubsection{Two-stage hybrid training}

Training variational quantum models directly on hardware is limited by latency and execution cost. To address this challenge, the framework adopts a two-stage hybrid training strategy that combines noise-aware simulation with hardware deployment.

\textbf{Phase 1: noise-aware simulation training.}
In the first phase, the hybrid model is trained on the BEAST GPU cluster using a noise-aware quantum simulator integrated into the PyTorch training loop. We use Qiskit Aer configured with the Emerald noise profile derived from calibration data of the physical processor. This lets the classical encoder--decoder networks and the quantum circuit parameters $\boldsymbol{\theta}$ converge under realistic noise conditions, while substantially reducing the cost of quantum hardware calls.

\textbf{Phase 2: hardware fine-tuning and inference.}
After convergence in simulation, the quantum backend is switched to the physical IQM Emerald processor through the IQM-Qiskit-provider. The structured ansatz uses only $N_q=9$ qubits, so multiple independent copies of the same circuit are placed in parallel on the $54$-qubit chip, yielding up to $4$ concurrent circuit executions per hardware call (due to the maintenance on several qubits); this parallelism is used in both the fine-tuning epochs that adapt $\boldsymbol{\theta}$ to hardware-specific noise and the final inference runs used for prediction and validation, which are executed directly on the quantum hardware.

All circuits are compiled and transpiled using Qiskit's transpile function targeting the Emerald backend. This process maps logical qubits to physical qubits, decomposes composite operations into native gates, optimises circuit depth, and inserts SWAP operations when required by the hardware connectivity. The native single-qubit gate of IQM Emerald is the parameterised $\mathrm{RX}$ pulse, which directly implements the mode-wise rotation $e^{-i\alpha_{\ell,i} X_i}$ of Eq.~\eqref{eq:Uq_circuit}, so the structured ansatz maps onto the native gate set without a basis change (Fig.~\ref{fig:qcml_device}b). Each hardware execution uses 4096 measurement shots, and expectation values are estimated by statistical averaging.

\subsubsection{Error mitigation}

Computation on NISQ devices is affected by multiple noise sources that can degrade measurement fidelity. We therefore apply measurement-error mitigation during the hardware fine-tuning and inference phase. The procedure calibrates the measurement confusion matrix of the relevant qubits and applies correction techniques such as matrix inversion or constrained least-squares reconstruction, implemented in libraries including mthree~\cite{nation2021scalable}. The output is a mitigated quasi-probability distribution used to estimate expectation values.

Depending on the dominant error channels observed on the device, additional mitigation strategies such as zero-noise extrapolation~\cite{he2020zero} or randomised compiling~\cite{wallman2016noise} can be added. In practice, the structured quantum circuits used in the framework are relatively shallow and carry few parameters, which further improves robustness to hardware noise compared with deeper generic variational circuits.

\bibliography{prex}

\section*{Acknowledgements}
P.V.C. and X.X. acknowledge funding support from the European Commission CompBioMed Centre of Excellence (Grant No. 675451 and 823712). Support from the UK Engineering and Physical Sciences Research Council under the following projects ``UK Consortium on Mesoscale Engineering Sciences (UKCOMES)'' (Grant No.EP/R029598/1) and ``Software Environment for Actionable and VVUQ-evaluated Exascale Applications (SEAVEA)'' (Grant No. EP/W007711/1) is gratefully acknowledged. P.V.C. and X.X. acknowledge the 2024-2025 DOE INCITE award for computational resources on supercomputers at the Oak Ridge Leadership Computing Facility under the ``COMPBIO3'' project. P.V.C. and X.X. acknowledge the use of resources provided by the Isambard-AI National AI Research Resource (AIRR). Isambard-AI is operated by the University of Bristol and is funded by the UK Government's Department for Science, Innovation and Technology (DSIT) via UK Research and Innovation; and the Science and Technology Facilities Council [ST/AIRR/I-A-I/1023]. We are grateful to IQM Quantum Computers for providing access to the Emerald superconducting quantum device. We thank Leibniz Supercomputing Centre (LRZ) for access to the BEAST GPU cluster.

\section*{Author contributions}
X.X. and M.W. conceptualised the approach. X.X. designed and ran the numerical simulations and generated the training and validation datasets. X.X. and M.W. developed the hybrid quantum–classical machine-learning framework and implemented the structured quantum latent propagator, including runs on the emulator and on the superconducting quantum device. M.G. performed the post-processing of the simulation data. M.C. contributed to the project discussions and planned access to BEAST at LRZ. P.V.C. supervised the project and provided access to the required infrastructure. All authors contributed to writing the manuscript.

\section*{Competing interests}
The authors declare no competing interests.


\clearpage

\end{document}